\documentclass{ws-p10x7}

\begin{document}

\title{Lattice Field Theory}

\author{Richard Kenway}

\address{Department of Physics and Astronomy, The University of
Edinburgh, Edinburgh EH9 3JZ, Scotland\\E-mail: r.d.kenway@ed.ac.uk}

\twocolumn[\maketitle\abstract{This review concentrates on progress in
lattice QCD during the last two years and, particularly, its impact on
phenomenology. The two main technical developments have been successful
implementations of lattice actions with exact chiral symmetry, and
results from simulations with two light dynamical flavours which provide
quantitative estimates of quenching effects for some quantities. Results
are presented for the hadron spectrum, quark masses, heavy-quark decays
and structure functions. Theoretical progress is encouraging renewed
attempts to compute non-leptonic kaon decays. Although computing power
continues to be a limitation, projects are underway to build
multi-teraflops machines over the next three years, which will be around
ten times more cost-effective than those of today.}]

\section{Introduction}

\subsection{Overview}

Progress in lattice QCD tends to be incremental. Through improved
formulations and increased computer power, we have been steadily
gaining control over all the systematic approximations inherent in
numerical simulations. Rarely, progress is revolutionary. Happily, we
are in the midst of such a major leap forward, through recent
demonstrations that lattice formulations, which preserve exact chiral
symmetry, work in practice. Combined with increasingly cost-effective
computing technology, a period of accelerated progress, impacting
directly on phenomenologically important QCD calculations, can be
foreseen over the next five years.

In this review, I describe recent progress demonstrating that chirally
symmetric formulations are feasible and results for phenomenologically
relevant quantities. I omit results for QCD thermodynamics and, with
one exception, for non-QCD theories. Although there has been a lot of
work on the challenging problem of the confinement mechanism, the
understanding achieved so far is partial and the picture is still too
confusing to do it justice in a review such as this. For results in
these areas, other exploratory phenomenological applications and most
of the technical details, I refer you to the latest in the annual
series of lattice conference proceedings\cite{lat99}.

\subsection{Objectives of Lattice Field Theory}

The primary objective of lattice field theory is to determine the
parameters of the Standard Model and, thereby, to seek signals of new
physics. Due to confinement, the quark sector is not directly
accessible by experiment and numerical simulations of QCD are needed
to provide the missing link. In principle, lattice QCD offers
model-independent computations of hadronic masses and matrix
elements. Ultimately, it should test QCD as the theory of strong
interactions and provide an understanding of confinement. The name of
the game is the control of systematics, particularly to quantify
dynamical quark effects and reliably simulate at, or extrapolate to
the physical values of the light and heavy quark masses.

The second objective is to determine the phase structure of hadronic
matter. Both the location and the order of the line of phase
transitions, separating the confined, hadronic phase from the
deconfined, quark-gluon plasma phase, in the temperature ($T$),
chemical potential ($\mu$) plane, are sensitive to the flavour
content. Since $T_c$ is close to the strange quark mass, it is
particularly important to simulate the strange quark accurately, and
this awaits simulations with more realistic dynamical light
flavours. Also, the $\mu\neq 0$ plane is not accessible to simulations
with current algorithms, because the action is complex and Monte Carlo
importance sampling fails. Significant progress has been made in QCD
thermodynamics at $\mu=0$, and new approaches using anisotropic
lattices should yield spectroscopic results to help better understand
leptonic decays of vector mesons, strangeness production and $J/\psi$
suppression.

Finally, looking beyond QCD, we aim to develop simulations into a
general purpose non-perturbative tool. Recent progress in formulating
lattice chiral symmetry has reawakened hopes of being able to simulate
chiral and SUSY theories.

\section{Theoretical Progress}

\subsection{Lattice Chiral Symmetry}

Major progress has been achieved over the past eight years following
the rediscovery of the Ginsparg-Wilson (GW)
relation\cite{gw_relation}. This states that if the lattice Dirac
operator, $D$, is chosen to satisfy
\begin{equation}
\gamma_5 D + D\gamma_5 = aD\gamma_5 D,
\label{eq:gw_relation}
\end{equation}
where $a$ is the lattice spacing, then the theory possesses an exact
chiral symmetry. Such a formulation has the great virtue that it
separates the chiral and continuum limits. It also forbids $O(a)$
terms, so that the resulting fermion actions are improved. The GW
relation languished for many years, because there was no practical
implementation. The breakthrough was the discovery of three
constructions: the overlap\cite{overlap}, domain
wall\cite{domain_wall} and perfect action\cite{perfect_action}. As a
consequence, lattice simulations with exact chiral symmetry are now a
reality (see Kikukawa's talk\cite{kikukawa} and other
reviews\cite{gw_reviews}).

For vector theories like QCD, this means that we can maintain a global
chiral symmetry at non-zero lattice spacing, so that simulations
should be able to approach the physical $u$ and $d$ quark masses in a
controlled fashion. More importantly, the mixing of operators of
different chiralities, which has plagued kaon mixing and decay
calculations, is avoided. Formally, abelian and non-abelian chiral
gauge theories have been constructed on the
lattice\cite{chiral_gauge_theories}, so the Standard Model can now be
defined non-perturbatively. Somewhat more speculatively, for SUSY
theories without scalar fields, lattice chiral symmetry forbids
relevant SUSY-violating terms in the action, and so offers the
prospect of lattice simulations without fine tuning\cite{lat0002030}.

\subsection{Overlap Quarks}

In the overlap formulation, the lattice Dirac operator has the form
\begin{equation}
D_{\rm ov}(\mu)=\left(\frac{1+\mu}{2}\right)
               +\left(\frac{1-\mu}{2}\right)\gamma_5{\rm sgn}(H_{\rm W}),
\label{eq:overlap}
\end{equation}
where $\mu$ is the bare quark mass, and $H_{\rm W}$ is the (hermitian)
Wilson-Dirac matrix with a large negative mass. $D_{\rm ov}(0)$ obeys
the GW relation (Eq.~(\ref{eq:gw_relation})). Numerically, the
challenge is to compute accurately the sign of the large sparse
matrix, $H_{\rm W}$, eg using a rational
approximation\cite{lat0001013}:
\begin{equation}
{\rm sgn}(H_{\rm W}) \approx \sum_{s=1}^N
                             \frac{1}{c_s^2+\frac{b_s^2}{H_{\rm W}}}.
\label{eq:approx_sgn}
\end{equation}
This approximation breaks down for small eigenvalues, so it is
necessary to project out the lowest eigenvectors and treat their signs
exactly.

\begin{figure*}
\vspace{-78mm}
\epsfxsize1.05\textwidth
\figurebox{}{}{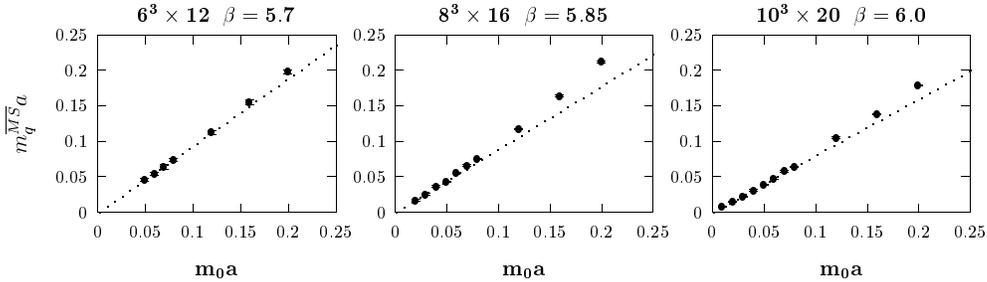}
\vspace{-85mm}
\caption{The renormalised quark mass, obtained from the axial Ward
identity, versus bare quark mass ($\mu=m_0a$) in quenched QCD,
computed using the overlap formulation at three different lattice
spacings\protect\cite{lat0006004}.}
\label{fig:overlap_quark_mass}
\end{figure*}

The resulting implementation has been tested for quenched
QCD\cite{lat0006004}. The results in
Figure~\ref{fig:overlap_quark_mass} show that, as expected, there is
no additive quark mass renormalisation, ie
\begin{equation}
m_q^{\overline{\rm MS}}a = Z_m^{-1}\mu\left[1+O(a^2)\right].
\label{eq:mass_ren}
\end{equation}
The quenched light hadron spectrum has also been
computed\cite{lat0006004}, albeit in a small volume, and evidence is
found that the mass ratios roughly scale at fixed quark mass, showing
that it is now possible to do phenomenologically interesting
calculations in these new chirally symmetric formulations.

Unfortunately, even for quenched QCD, the computational cost of
simulating overlap quarks is comparable to that of dynamical
simulations\cite{lat0001008}, due to the extra work involved in
computing the sign function (Eq.~(\ref{eq:approx_sgn})). Little is
known yet about the cost of simulating dynamical quarks this way.

The approximation of the sign function in Eq.~(\ref{eq:approx_sgn})
may be replaced by local interations amongst a set of $2N$ auxiliary
fields:
\begin{eqnarray}
\bar{\psi}&{\rm sgn}(H_{\rm W})&\psi
          \approx \bar{\psi}\sum_{s=1}^N
          \frac{1}{c_s^2+\frac{b_s^2}{H_{\rm W}}}\psi\nonumber\\
& \rightarrow & \hspace{-1em}\sum_s
          \left[(\bar{\psi}_s\chi_s+\bar{\chi}_s\psi_s) 
          + \bar{\chi}_s(c_s^2H_{\rm W})\chi_s\right.\nonumber\\
&&\hspace{-1em}\left.+ b_s(\bar{\chi}_s\phi_s+\bar{\phi}_s\chi_s) 
          - \bar{\phi}_sH_{\rm W}\phi_s\right].
\label{eq:5d_overlap}
\end{eqnarray}
This shows that the overlap formulation may be thought of as
five-dimensional, in which the fifth dimension is like
flavour\cite{lat0005004}.

\subsection{Domain Wall Quarks}

Here the fermions live in five dimensions and are coupled to a mass
defect located on a four-dimensional hyperplane. On a finite lattice,
with $L_s$ sites in the fifth dimension, the Dirac operator may be
written
\begin{eqnarray}
D_{\rm DW}(\mu) &=& \left(\frac{1+\mu}{2}\right)
                    +\left(\frac{1-\mu}{2}\right)\nonumber\\
                 && \times\gamma_5\tanh\left(-\frac{L_s}{2}\log T\right).
\label{eq:domain_wall_operator}
\end{eqnarray}
Here $T$ is the transfer matrix in the fifth dimension and $\mu$ is
again the bare mass. The strong similarity with the overlap,
Eq.~(\ref{eq:overlap}), is evident. In fact, the two formulations are
identical in the limits of $L_s\rightarrow\infty$ and zero lattice
spacing in the fifth direction.

For finite $L_s$, the chiral modes of opposite chirality are trapped
on the four-dimensional domain walls at each end of the fifth
dimension. Chiral symmetry is broken, but the breaking is
exponentially suppressed by the size of the fifth dimension. Several
groups have tested this in quenched
QCD\cite{lat9909117,lat0007014}. It is found that the pion mass does
not always vanish with quark mass, in the limit
$L_s\rightarrow\infty$, due to near unit eigenvalues of $T$, which
allow unsuppressed interactions between the LH and RH fermions. This
is a strong-coupling effect, which goes away for weak enough coupling,
or using a renormalisation-group improved action\cite{lat0007014}. The
problem may be controlled numerically by projecting out the low
eigenvectors and taking their contribution with infinite
$L_s$\cite{lat0005002}.

Thus, there is now a good understanding of how to achieve a close
approximation to exact chiral symmetry in lattice QCD
simulations. Although numerically relatively expensive, it is early
days and more efficient algorithms may yet be found. Already, the
extra degree of control given by exact symmetry probably outweighs the
cost for matrix element calculations, and we can anticipate rapid
progress over the next few years.

\section{String Breaking}

For the remainder of this review, I will describe results obtained
using traditional formulations of lattice QCD, focusing on the effects
of dynamical quarks in order to try to quantify quenching errors for
as many quantities as possible.

\begin{figure}
\vspace{-3mm}
\epsfxsize0.9\columnwidth
\figurebox{}{}{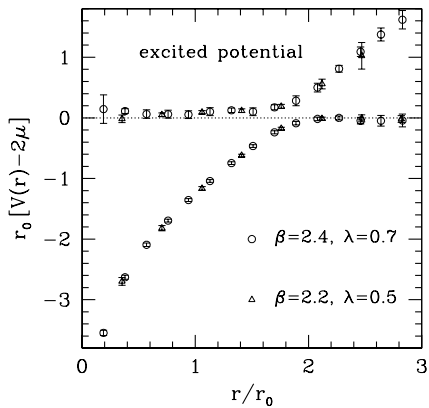}
\vspace{-5mm}
\caption{String breaking in the four-dimensional SU(2) Higgs
model\protect\cite{lat9909164}. The results for the potential, $V(r)$,
versus separation, $r$, at two lattice spacings are in excellent
agreement.}
\label{fig:string_breaking}
\end{figure}

Perhaps, the first effect of dynamical quarks we might hope to see is
string breaking. This flattening of the potential between two static
charges, as they are separated, has been observed as level crossing in
the confinement phase of the SU(2) Higgs model\cite{lat9909164}. At a
particular separation, the string state becomes degenerate with the
state of two static-light ``mesons'', as shown in
Figure~\ref{fig:string_breaking}.

Demonstrating string breaking in QCD at zero temperature is proving to
be much more challenging than expected. As yet there is no completely
convincing signal. This is because there is poor overlap between the
string states used to compute the static quark potential and the
broken-string state, comprising two static-light mesons. Including the
latter is computationally very costly, because it requires quark
propagators at all sites.

However, the mixing matrix element has been computed for two dynamical
flavours and found to be non-zero\cite{lat0001015}. Using only string
states, there are hints of a flattening potential, just as the signal
becomes swamped by noise\cite{lat9909118}, but a recent
high-statistics calculation provides pretty conclusive evidence that
such attempts are doomed\cite{lat0005018} and the two-meson state must
be included (as was done for the Higgs model\cite{lat9909164}). The
situation is frustrating, but it is only a matter of time before
sufficient computing resources are brought to bear.

\section{Hadron Spectrum}

The important result that the quenched light hadron spectrum disagrees
with experiment was finally established by the CP-PACS
Collaboration\cite{lat9904012} in 1998 and announced at ICHEP98. This
had proved difficult, requiring high statistics, because the deviation
is less than 10\%. This small deviation is good news for phenomenological
applications of lattice QCD, which still rely heavily on the quenched
approximation. The main symptom of quenching is that it is not
possible consistently to define the strange quark mass -- the two
spectra obtained from using the $K$ and the $\phi$ meson to determine
the strange quark mass disagree.

\begin{figure}
\epsfxsize0.85\columnwidth
\figurebox{}{}{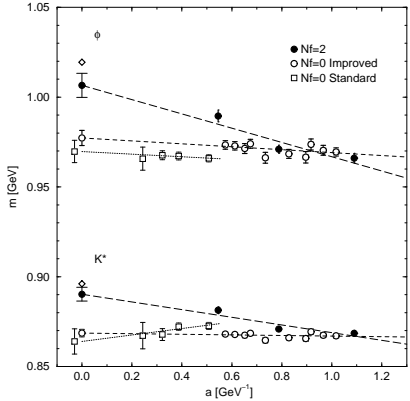}
\vspace{-8mm}
\caption{Comparison of continuum extrapolations of the $K^*$ and
$\phi$ meson masses in two-flavour (filled symbols) and quenched (open
symbols) QCD\protect\cite{kanaya}. The $K$ mass was used to fix the
strange quark mass.}
\label{fig:strange_spectrum}
\end{figure}

Since 1998, the focus has been on simulations with two degenerate
dynamical flavours, which are identified with the $u$ and $d$
quarks. The strange and higher-mass quarks are still treated in the
quenched approximation. At this conference, CP-PACS reported that the
resulting strange meson spectrum is much closer to
experiment\cite{kanaya}, as shown in
Figure~\ref{fig:strange_spectrum}.

The glueball spectrum has only been computed in quenched QCD and the
results\cite{quenched_glueballs} reported at ICHEP98 remain state of
the art. This calculation was hard because of strong scaling
violations. Better scaling has been reported recently using the
fixed-point action\cite{lat0007007}. Mixing with quark states should
be very important. A first attempt to compute the mixing of the
lightest scalar glueball with the lightest scalar quarkonium states
has concluded that the $f_0(1710)$ is 74\% glueball, whereas the
$f_0(1500)$ is 98\% quarkonium\cite{lat9910008}.

Now that two-flavour simulations are possible, it is interesting to
try to compute the flavour-singlet meson masses. A first attempt has
produced a result that the $\eta$, $\eta'$ mixing angle is around
$45^\circ$ in the $\{(\bar{u}u+\bar{d}d)/\sqrt{2}, \bar{s}s\}$ basis,
but, despite sophisticated variance reduction techniques, at least
ten-times better statistics is needed\cite{lat0006020}.

\section{Quark Masses}

Quark masses are encoded within hadron masses. They are scale and
renormalisation-scheme dependent. To be useful for phenomenology, it
is necessary to compute the scale evolution of the quark mass, from
the lattice scale at which it is determined, to a suitably high scale
where it can be matched to a perturbative scheme.

The usual approach is to define an intermediate scheme, such as the
Schr\"odinger Functional (SF), in which the scale dependence can be
computed non-perturbatively (see Heitger's talk\cite{heitger}). Once
this scale dependence is known to a high enough energy, it can be
continued to infinite energy, using the perturbative renormalisation
group, to define the ratio of the lattice mass to the
renormalisation-group (RG) invariant mass,
$M$\cite{alpha_quark_mass_scaling}. Thus the lattice quark mass fixes
$M$ and, from it, perturbation theory may be used to determine the
quark mass in any chosen scheme.

\begin{figure}
\vspace{-42mm}
\epsfxsize1.1\columnwidth
\figurebox{}{}{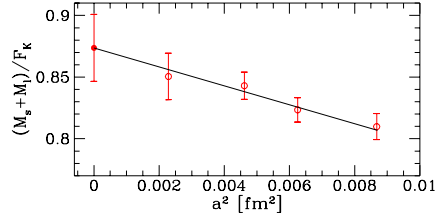}
\vspace{-10mm}
\caption{Continuum extrapolation of the RG-invariant mass, $M_s+M_l$,
where $M_l = (M_u+M_d)/2$, in units of the $K$ decay constant for
quenched QCD\protect\cite{lat9906013}.}
\label{fig:quenched_strange_mass}
\end{figure}

Figure~\ref{fig:quenched_strange_mass} shows the sum of the
RG-invariant average $u$ and $d$, and $s$ quark masses, computed in
this way for quenched QCD (with the $K$ mass as
input)\cite{lat9906013}. The resulting estimate for the strange quark
mass is
\begin{equation}
m_s^{\overline{\rm MS}}(2\;{\rm GeV})=97(4)\;{\rm MeV}\hspace{1em}(N_f=0).
\label{eq:quenched_strange_mass}
\end{equation}

\begin{figure}
\vspace{-5mm}
\epsfxsize\columnwidth
\figurebox{}{}{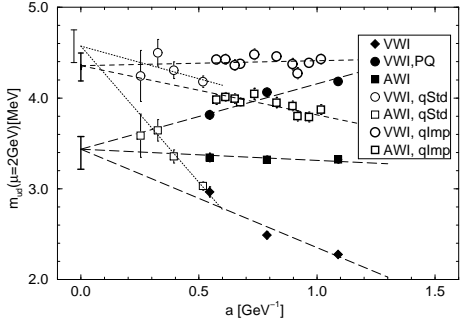}
\vspace{-10mm}
\caption{Continuum extrapolation of the average $u$ and $d$ quark
masses for $N_f=2$ (filled symbols) and $N_f=0$ (open symbols)
QCD\protect\cite{kanaya,lat0004010}.}
\label{fig:ud_quark_mass}
\end{figure}

At this conference, CP-PACS announced results for light quark masses
in two-flavour QCD\cite{kanaya}, updating earlier
results\cite{lat0004010}. They use an improved action, with a fixed
lattice size of $(2.5~{\rm fm})^3$, and extrapolate downwards in the
sea-quark mass, from masses corresponding to pseudoscalar-to-vector
meson mass ratios above 0.6, to the average $u$ and $d$ quark
mass. Currently, non-perturbative matching is not available for
two-flavour QCD, so CP-PACS uses mean-field improved 1-loop
matching. The results are shown in Figure~\ref{fig:ud_quark_mass}, for
three different definitions of quark mass at non-zero lattice spacing,
and compared with quenched QCD. The different definitions permit
consistent continuum extrapolations and the final results are from
combined fits with a single limit.  They find a big effect from the
inclusion of dynamical quarks. The average $u$ and $d$ quark mass is
reduced by roughly 25\%:
\begin{eqnarray}
m_{ud}^{\overline{\rm MS}}(2{\rm GeV}) 
&=& 3.44^{+0.14}_{-0.22}{\rm MeV}(N_f\hspace{-1mm}=2)\\
m_{ud}^{\overline{\rm MS}}(2{\rm GeV}) 
&=& 4.36^{+0.14}_{-0.17}{\rm MeV}(N_f\hspace{-1mm}=0).
\label{eq:ud_quark_mass}
\end{eqnarray}

\begin{figure}
\vspace{-5mm}
\epsfxsize\columnwidth
\figurebox{}{}{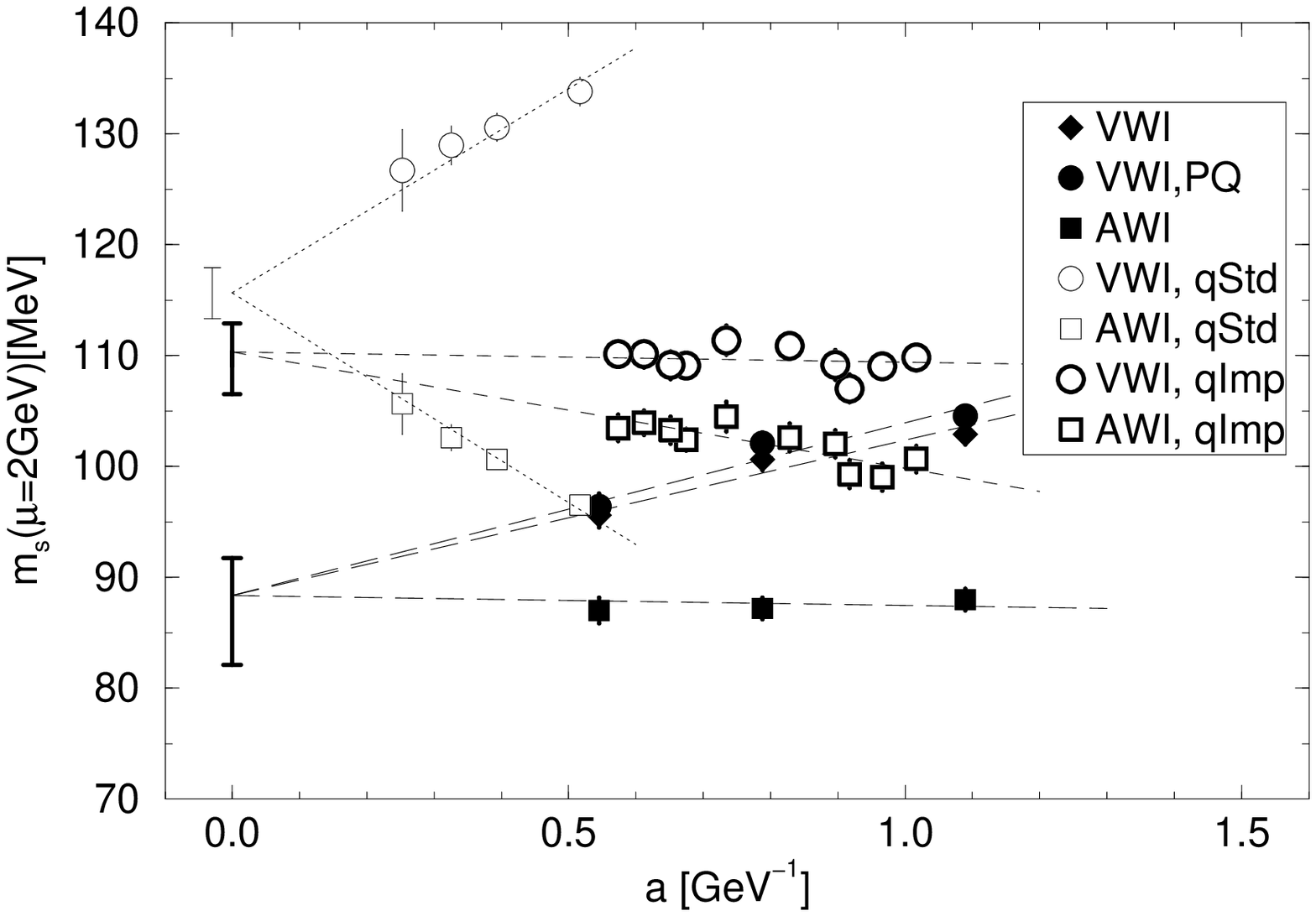}\\[-9mm]
\epsfxsize\columnwidth
\figurebox{}{}{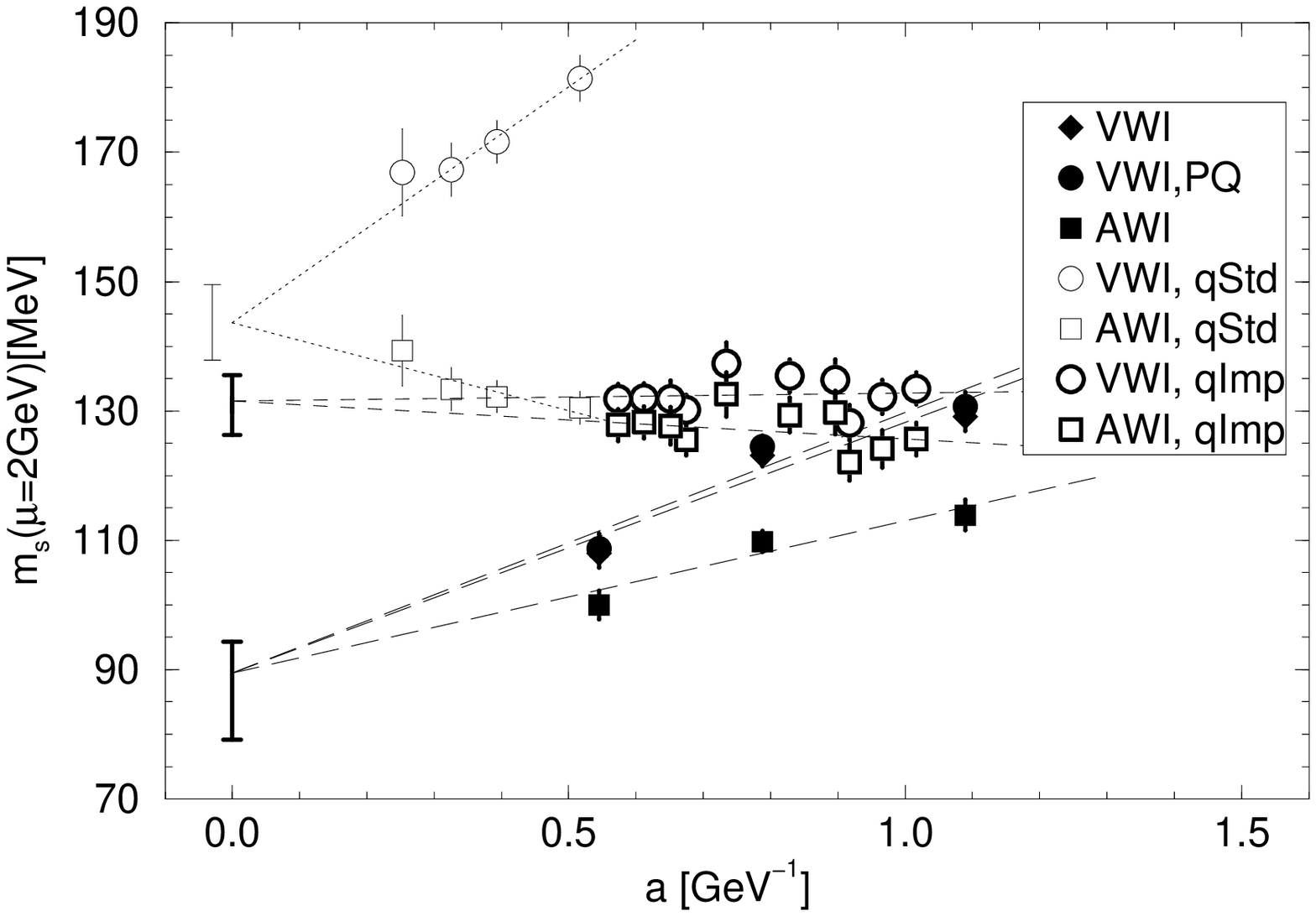}
\vspace{-10mm}
\caption{Continuum extrapolation of the strange quark mass for $N_f=2$
(filled symbols) and $N_f=0$ (open symbols) QCD, determined from the
$K$ mass (upper figure) and from the $\phi$ mass
(lower figure)\protect\cite{kanaya,lat0004010}.}
\label{fig:s_quark_mass}
\end{figure}

Treating the strange quark in the quenched approximation, CP-PACS
finds that the 20\% inconsistency in the strange quark mass in
quenched QCD disappears with dynamical $u$ and $d$ quarks (within 10\%
errors), as can be seen in Figure~\ref{fig:s_quark_mass}. The
continuum estimates are
\begin{eqnarray}
m_{s}^{\overline{\rm MS}}(2{\rm GeV}) 
&=& 88^{+4}_{-6}{\rm MeV}\mbox{ ($K$ mass)}\\
m_{s}^{\overline{\rm MS}}(2{\rm GeV}) 
&=& 90^{+5}_{-11}{\rm MeV}\mbox{ ($\phi$ mass)}.
\label{eq:s_quark_mass}
\end{eqnarray}
Again, the mass is reduced substantially compared to quenched
QCD. Such a low strange quark mass suggests a large value of
$\epsilon'/\epsilon$, and raises the interesting question how much
lower the strange quark mass would be if it were treated
dynamically. The result $m_s/m_{ud}=26(2)$\cite{kanaya} agrees with
the chiral perturbation theory estimate of 24.4(1.5).

The $b$ quark mass, in this world in which only the $u$ and $d$ quarks
are dynamical, obtained from the $B_s$ binding energy at leading order
in $1/m_b$ and using NNLO perturbative matching, is\cite{lat0002007}
\begin{equation}
m_b^{\overline{\rm MS}}(m_b^{\overline{\rm MS}})=4.26(9)~{\rm GeV}.
\label{eq:b_mass}
\end{equation}

\section{Heavy-Quark Decays}

The calculation of hadronic matrix elements associated with the weak
decays of $b$ quarks is the most successful phenomenological
application of lattice QCD (see Kronfeld's talk\cite{kronfeld}).

\subsection{Leptonic Decays and Mixing}

The top-quark CKM matrix elements and the neutral $B_q$ mass
difference are related through the hadronic matrix element
$f_{B_q}\sqrt{\hat{B}_{B_q}}$:
\begin{eqnarray}
\Delta m_q &=& 
           \frac{G_F^2}{6\pi^2}M_W^2\eta_BS_0(m_t^2/M_W^2)M_{B_q}\nonumber\\
           && \hspace{10mm}\times|V_{tq}V_{tb}^\ast|^2f_{B_q}^2\hat{B}_{B_q}.
\label{eq:mixing}
\end{eqnarray}
Traditionally, $f_B$ and $\hat{B}_B$ have been computed separately in
lattice QCD. Quenched estimates for $f_B$ have stabilised in recent
years, but suffer a relatively large irreducible scale uncertainty due
to the quenched approximation. Including two dynamical flavours
increases $f_B$ by around 20\%, although statistical errors currently
overwhelm systematic effects. Table~\ref{tab:f_B} shows the best
estimates of the $B$ decay constants, presented at Lattice 99. The
only direct comparison with experiment is for $f_{D_s}$, and here the
lattice estimates,
\begin{eqnarray}
f_{D_s} &=& 241(30)~{\rm MeV}~(N_f=0)\mbox{\cite{lat0007020}}\\
f_{D_s} &=& 275 (20)~{\rm MeV}~(N_f=2)\mbox{\cite{kanaya}},
\label{eq:f_Ds}
\end{eqnarray}
are consistent with experiment (eg, ALEPH's result of
285(45)~MeV\cite{golutvin}), although the errors are too large to
expose systematic effects.

\begin{table}
\caption{Summary of lattice results for $B$ decay constants, presented
by Hashimoto at Lattice 99\protect\cite{lat9909136}.}
\label{tab:f_B}
\begin{center}
\begin{tabular}{|l|c|c|} 
\hline 
                & $N_f=2$ & $N_f=0$\\
\hline
$f_B$~(MeV)     & 210(30) & 170(20)\\
$f_{B_s}$~(MeV) & 245(30) & 195(20)\\
$f_{B_s}/f_B$   & 1.16(4) & 1.15(4)\\
\hline
\end{tabular}
\end{center}
\end{table}

The combination $f_B\sqrt{\hat{B}_B}$ may be computed directly in
lattice QCD and, to the extent that systematic errors in $f_B$ and
$\hat{B}_B$ are correlated, this may be more reliable than separate
determinations of $f_B$ and $\hat{B}_B$. A recent
non-perturbatively-renormalised result in quenched QCD
is\cite{lat0002025}
\begin{equation}
f_B\sqrt{\hat{B}_B} = 206(29)~{\rm MeV}.
\label{eq:f_B_root_B_B}
\end{equation}
Systematics should also cancel in ratios, so that quenched results
such as\cite{lat0002025}
\begin{eqnarray}
\frac{f_B}{f_{D_s}} &=& 0.74(5)\\
\frac{f_{B_s}\sqrt{\hat{B}_{B_s}}}{f_B\sqrt{\hat{B}_B}} &=& 1.16(7)
\label{eq:ratios_of_decay_const}
\end{eqnarray}
are probably the most reliable.

\subsection{Lifetimes}

Lifetime calculations are at an exploratory stage. Two groups have
computed the $B_s$ lifetime difference,
$\Delta\Gamma_{B_s}/\Gamma_{B_s}$, obtaining
0.047(15)(16)\cite{ph0006135} and 0.107(26)(14)(17)\cite{lat0004022},
to be compared with the experimental upper bound of 0.31. Although
they use quite different lattice techniques, the matrix element
calculations are consistent, and the discrepancy in the final results
is due to one using the quenched, and the other using the unquenched
value of $f_{B_s}$.

The $\Lambda_b$ lifetime is a puzzle, because the experimental
measurement for the ratio $\tau(\Lambda_b)/\tau(B_0)=0.79(5)$ is
significantly different from one, whereas, to leading order in the
heavy-quark mass, all $b$ hadrons have the same lifetime. A
preliminary lattice calculation, however, does indicate that spectator
effects are significant at the 6--10\% level\cite{lat9906031}.

\subsection{Exclusive Semileptonic Decays}

The main purpose of computing exclusive semileptonic $B$ decay form
factors is to extract model-independent estimates of $|V_{ub}|$ and
$|V_{cb}|$ from experiment. All the results I present were computed in
quenched QCD, although dynamical-quark effects are now beginning to be
explored\cite{lat9909076}. They are expected to be around 10\%.

\begin{figure}
\vspace{-13mm}
\epsfxsize1.05\columnwidth
\figurebox{}{}{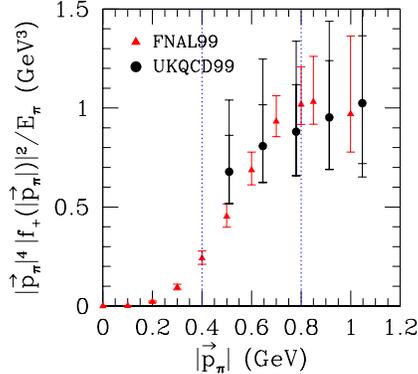}
\vspace{-12mm}
\caption{Results for the $B\rightarrow\pi\ell\nu$ differential decay
rate, compiled by Lellouch\protect\cite{ph9912353}, from
FNAL\protect\cite{lat9910010} and
UKQCD\protect\cite{lat9911011}. Vertical lines indicate the momentum
range, 0.4~GeV$\leq|\vec{p}_\pi|\leq$ 0.8~GeV, within which the
lattice artefacts are minimised and comparison with experiment should
be most reliable\protect\cite{lat9910010}.}
\label{fig:B_to_pi_diff_rate}
\end{figure}

During the past two years, the most progress has been in calculating
the form factors for $B\rightarrow\pi\ell\nu$, defined by
\begin{eqnarray}
\langle\pi(p')|V^\mu|B(p)\rangle 
&=& \frac{M_B^2-M_\pi^2}{q^2}q^\mu f_0(q^2)\nonumber\\
&& \hspace{-28mm} + \left(p^\mu+p'^\mu
   -\frac{M_B^2-M_\pi^2}{q^2}q^\mu\right)f_+(q^2)\\
q=p-p', && V^\mu=\bar{b}\gamma^\mu u.
\label{eq:B_to_pi_form_factors}
\end{eqnarray}
Here, lattice QCD fixes the normalisation of the form factors, unlike
heavy-quark effective theory (HQET). However, today's lattice spacings
are too large to represent a high-momentum pion accurately. So the
kinematic range is restricted to near zero recoil, and
model-independent results are only possible for the differential decay
rate\cite{lat9910010,lat9911011}, as shown in
Figure~\ref{fig:B_to_pi_diff_rate}. Differential rates should be
measured experimentally soon, at which point, direct comparison with
the lattice results will provide a model-independent estimate for
$|V_{ub}|$.

\begin{figure}
\vspace{-13mm}
\epsfxsize1.05\columnwidth
\figurebox{}{}{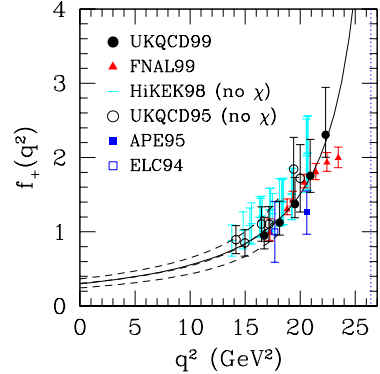}
\vspace{-12mm}
\caption{Results from various lattice groups for the
$B\rightarrow\pi\ell\nu$ form factor $f_+$, compiled by
Lellouch\protect\cite{ph9912353}. The solid line is a fit to the UKQCD
data and the dashed lines are light-cone sum rule results.}
\label{fig:B_to_pi_f_+}
\end{figure}

Model-dependent extrapolation is needed to obtain the full kinematic
range. A dipole fit to the UKQCD results for $f_+(q^2)$ is shown in
Figure~\ref{fig:B_to_pi_f_+} (a simultaneous pole fit to $f_0(q^2)$
imposes the constraint $f_0(0)=f_+(0)$). This fit gives a total decay
rate
\begin{equation}
\Gamma/|V_{ub}|^2 = 9^{+3+2}_{-2-2}~{\rm ps}^{-1}.
\end{equation}
There is good agreement with the results from other groups and with
light-cone sum rules at low $q^2$. The form factor $f_0$ provides an
important consistency check on the lattice results through the
soft-pion theorem:
\begin{equation}
f_0(q^2_{\rm max})=f_B/f_\pi
\label{eq:soft_pion}
\end{equation}
in the limit of zero pion mass. Whether this is satisfied by current
simulations is somewhat controversial\cite{lat9909100,lat9909076}, but
this will have to be resolved if we are to have full confidence in the
lattice results.

There have been no new results for $B\rightarrow\rho\ell\nu$ and
$B\rightarrow K^\ast\gamma$ and the present status is described in
Lellouch's review\cite{ph9912353}.

The form factors for the heavy-to-heavy decays, $B\rightarrow
D^{(\ast)}\ell\nu$, are better suited to lattice calculations, because
the recoil is smaller and present-day lattices can cover the full
kinematic range. However, HQET is able to determine the normalisation
at zero recoil in the heavy-quark symmetry limit. So, lattice QCD is
left with the tougher task of quantifying the deviations from the
symmetry limit at physical quark masses, needed to extract
$|V_{cb}|$ from experiment, which requires few percent accuracy. The
FNAL group has devised a technique for determining these power
corrections at zero recoil, from ratios of matrix elements, in which
many statistical and systematic errors
cancel\cite{kronfeld,ph9906376}, obtaining, eg,
\begin{equation}
F_{D^\ast}(1) = 0.935(22)(^8_{11})(8)(20).
\label{eq:zero_recoil}
\end{equation}

\begin{figure}
\vspace{-3mm}
\epsfxsize0.8\columnwidth
\figurebox{}{}{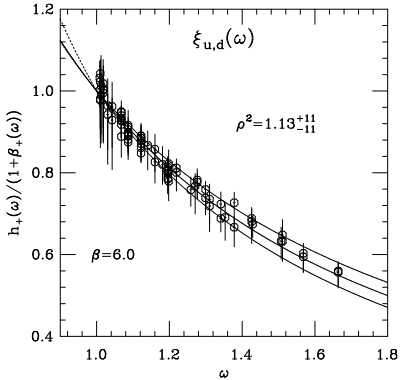}
\vspace{-4mm}
\caption{The Isgur-Wise function for quenched
QCD\protect\cite{lat9909126}, as a function of the recoil
$\omega$. $\rho$ is the slope parameter at $\omega=1$.}
\label{fig:isgur_wise}
\end{figure}

From the recoil dependence of the form factors, it is possible to
extract the Isgur-Wise function, $\xi(\omega)$. UKQCD has done this in
quenched QCD\cite{lat9909126}, using the $B\rightarrow D\ell\nu$ form
factor. The resulting estimate of $\xi(\omega)$, shown in
Figure~\ref{fig:isgur_wise}, is independent of the heavy-quark masses,
for masses around that of the charm quark, and is insensitive to the
lattice spacings used. Thus, it is demonstrably the Isgur-Wise
function for quenched QCD.

\section{Kaon Physics}

\subsection{Mixing}

The $B$-parameter for neutral kaon mixing,
\begin{equation}
B_K=\frac{3}{8}\frac{\langle\overline{K}^0|\bar{s}\gamma_\mu(1-\gamma_5)d
                     \bar{s}\gamma_\mu(1-\gamma_5)d|K^0\rangle}
                     {\langle\overline{K}^0|\bar{s}\gamma_\mu\gamma_5d|0\rangle
                     \langle 0|\bar{s}\gamma_\mu\gamma_5d|K^0\rangle},
\label{eq:B_K}
\end{equation}
is probably the best determined weak matrix element in quenched
QCD. The most reliable result comes from staggered quarks, because the
mixing due to chiral symmetry breaking is non-trivial for Wilson
quarks. The error in the continuum result\cite{lat9910032},
\begin{equation}
B_K(2~{\rm GeV}) = 0.628(42),
\label{eq:B_K_result}
\end{equation}
is mostly from perturbative matching, and should be reduced when
non-perturbative renormalisation becomes available. Dynamical quark
effects raise $B_K$ by around 5\% at fixed lattice spacing, but it is
not known yet how this affects the continuum result.

\subsection{Non-Leptonic Decays}

Although there are no new results, there is renewed optimism for
lattice calculations of $K\rightarrow\pi\pi$ decays (see Testa's
talk\cite{testa}), because we now have more sophisticated techniques,
which may afford control over the severe cancellations between the
matrix elements concerned. The new chirally symmetric lattice
formulations should avoid the mixing between operators of different
chiralities, and the large measured value for $\epsilon'/\epsilon$
reassures us that a signal should exist.
 
The fundamental problem for lattice QCD is that, according to the
Maiani-Testa no-go theorem, there is no general method for dealing
with multi-hadron final states in Euclidean space. The traditional
approach around this is to use chiral perturbation theory to relate
those matrix elements which can be computed, such as $K\rightarrow$
vacuum, $\pi$, or $\pi\pi$ at unphysical momenta, to the desired
physical matrix elements\cite{lat0006029}. A new proposal is to tune
the lattice volume so that one of the (discrete) energy levels of the
two pions equals the $K$ mass, and then relate the transition matrix
element to the decay rate in infinite volume\cite{lat0003023}.

The impending flood of experimental data for non-leptonic $B$ decays
presents a formidable challenge to lattice QCD. Chiral perturbation
theory no longer helps. Perhaps, the $B\rightarrow\pi\pi$
factorisation\cite{ph0006124}, proved for $m_b\gg\Lambda_{\rm QCD}$,
can be exploited in some way?

\section{Structure Functions}

Lattice QCD can provide the normalisation for parton
densities. Ultimately, this should test QCD and the validity of
perturbation theory. It enables us to disentangle power corrections
and, where experimental information is scarce, such as for the gluon
distribution for $x>0.4$, lattice QCD can help
phenomenology. Dynamical quark effects are presumably
crucial. Although results so far are for quenched QCD, this will soon
change (see Jansen's talk\cite{jansen}).

The traditional approach uses the operator product expansion to relate
moments of structure functions to hadronic matrix elements of local
operators:
\begin{eqnarray}
{\cal M}_n(q^2) &=& \int_0^1 {\rm d}x\;x^{n-2}F_2(x,q^2)\nonumber\\
&=& C_n^{(2)}(q^2/\mu^2,g(\mu))A_n^{(2)}(\mu)\nonumber\\
&& +\; O(1/q^2).
\label{eq:ope}
\end{eqnarray}
The Wilson coefficients, $C_n^{(2)}$, are determined in perturbation
theory and the hadronic matrix elements, $A_n^{(2)}$ are determined
on the lattice. Renormalisation is the major source of systematic
error, since the product $C(\mu)A(\mu)$ must be independent of the
scale $\mu$. This is achieved using a non-perturbative intermediate
scheme, in the same way as for quark masses:
\begin{equation}
A(\mu) = Z_{\rm INT}^{\overline{\rm MS}}(\mu)Z_{\rm latt}^{\rm INT}(\mu a)
         A^{\rm latt}(a).
\label{eq:renorm}
\end{equation}
The INT=SF scheme\cite{lat9901016} uses a step scaling function to
relate the matrix element, renormalised at the lattice scale, to a
high scale where perturbation theory can be used to determine the
RG-invariant matrix element in the limit $\mu\rightarrow\infty$. This,
in turn, may be related via perturbation theory to $\overline{\rm
MS}$. At this conference, Jansen\cite{jansen} reported that the
average momentum of partons in the pion in quenched QCD, computed in
this way, is
\begin{equation}
\langle x\rangle(2.4~{\rm GeV}) = 0.30(3),
\label{eq:ave_x}
\end{equation}
to be compared with the experimental result of 0.23(2), and confirming
early lattice results that the quenched estimate is larger than
experiment.

\begin{figure}
\vspace{-3mm}
\epsfxsize0.75\columnwidth
\figurebox{}{}{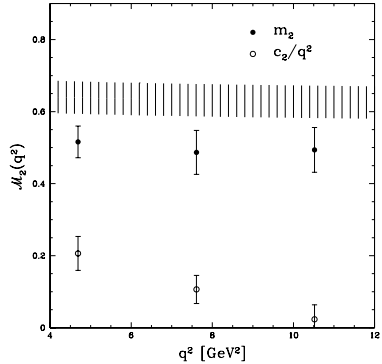}
\caption{The leading contribution, $m_2$, and the power correction,
$c_2/q^2$, to the lowest non-trivial moment of the unpolarised nucleon
structure function in quenched QCD\protect\cite{ph9906320}. The
hatched strip is the result without the inclusion of higher-twist
effects (the width of the strip indicates the error).}
\label{fig:m_2}
\end{figure}

A quite different method involves computing the current-current matrix
element, $\langle h|J_\mu J_\nu|h\rangle$, which appears in the
cross-section, directly on the lattice\cite{ph9906320}. The Wilson
coefficients are determined non-perturbatively from matrix elements
between quark states, by inverting
\begin{eqnarray}
\langle p|J_\mu(q)J_\nu(-q)|p\rangle 
&=& \sum_{m,n}C^{(m)}_{\mu\nu,\mu_1\ldots\mu_n}(a,q)\nonumber\\
&& \hspace{-3mm}\times\langle p|O^{(m)}_{\mu_1\ldots\mu_n}(a)|p\rangle,
\label{eq:non_pert_ope}
\end{eqnarray}
thereby avoiding mixing and renormalon ambiguities. Using 62 operators
and 70 momenta to extract the $C$'s, and reconstructing $\langle
N|J_\mu J_\nu|N\rangle$ from them and nucleon matrix elements, QCDSF
obtained the lowest non-trivial moment of the unpolarised structure
function\cite{ph9906320}, shown in Figure~\ref{fig:m_2}. This
indicates large power corrections and strong mixing between twist-2
and twist-4 operators.

\section{Machines and Prospects}

Progress in lattice QCD, and particularly its application to
phenomenology, continues to be critically dependent on increasing
computer power. Three machines dominate lattice QCD
today. Historically, the first was CP-PACS's 300~Gflops (sustained)
Hitachi SR2201. This has been operating since 1996 and cost
approximately \$70/Mflops. The second was the QCDSP, custom built
using 32-bit digital signal processors, which has been sustaining
120~Gflops and 180~Gflops at Columbia and Brookhaven, respectively,
since 1998. Its cost was around \$10/Mflops. This year, APE's latest
fully-customised 32-bit machine, called APEmille, began operation at
Pisa, Rome and Zeuthen, sustaining around 70~Gflops in the largest
configuration so far (this will double by the end of 2000). Its cost
is \$5/Mflops. These machines show an encouraging trend towards
greater cost-effectiveness.

In December 1999, an ECFA Working Panel concluded\cite{ecfa} that
``the future research programme using lattice simulations is a very
rich one, investigating problems of central importance to the
development of our understanding of particle physics''. It also
concluded that ``to remain competitive, the community will require
access to a number of 10 Tflops machines by 2003'' and ``it is
unlikely to be able to procure a 10 Tflops machine commercially at a
reasonable price by 2003''. 

Two new projects are targeting 10~Tflops 64-bit machines, with a
price/performance of \$1/Mflops, by 2003. The QCDOC project, involving
Columbia and UKQCD will employ PowerPC nodes in a 4-dimensional mesh
interconnect. The apeNEXT project, involving INFN and DESY, will
continue the APE architecture of custom nodes in a 3-dimensional
mesh. Two US projects, Cornell-Fermilab-MILC and JLAB-MIT, are
exploring Alpha and Pentium clusters using commodity (Myrinet)
interconnect, in the hope that these commodity components can be made
to scale to many thousands of processors and that the intense market
competition will drive the price very low. These developments,
together with the highly parallel algorithms employed for QCD, suggest
there will be no obstacle to multi-teraflops machines for QCD except
money!

However, we still do not understand the scaling of our algorithms well
enough to predict how much computer power will be needed. Our best
estimates are that to achieve comparable precision to quenched QCD, in
simulations with two dynamical flavours with masses around 15~MeV,
will require between 15 and 150 Tflops
years\cite{lat9911016}. However, we know nothing about simulations
with light enough quarks for $\rho\rightarrow\pi\pi$! We still have a
great deal to learn.

In conclusion, the range of phenomenological applications of lattice
QCD continues to expand. Key developments have been improved actions
(reported at ICHEP98) and non-perturbative renormalisation, both of
which have considerably increased our confidence in matrix element
calculations. Lattice QCD continues to drive the development of
cost-effective high-performance computing technology and there is no
technological limit in sight. The primary objective will be to extend
the range of quark masses which may be simulated reliably, and it is
hard to see how we will get away with less than 100~Tflops
machines. Finally, in the discovery of lattice chiral symmetry, we are
witnessing the ``Second Lattice Field Theory Revolution'', and this
will vastly increase the reach of {\it ab initio} computer
simulations.

\end{document}